**On the accuracy of measurements of electron temperature by Thomson scattering diagnostic in the plasma core of the ITER tokamak**

**M.Yu. Kantor**

**Ioffe Institute, Saint-Petersburg, Russia**

## Abstract

The ITER tokamak project includes a Thomson scattering diagnostic designed to measure electron temperature and density in the plasma core. The system is required to provide measurements over a wide temperature range while meeting stringent accuracy requirements. A previous study analyzed the errors of electron temperature measurements to assess the feasibility of these requirements. The analysis concluded that the central electron temperature could be measured with the required accuracy of 10% for temperatures up to 40 keV at a minimum electron density of $3\times10^{19}$ m$^{-3}$. However, those results were based on an overestimation of the number of photoelectrons generated in detectors by the scattered radiation due to the incorrect application of the Thomson scattering cross-section. As a consequence, the temperature measurement errors were significantly underestimated. In the present work, the accuracy of electron temperature measurements in the ITER plasma core is reassessed using the corrected photoelectron yield derived from published data and incorporating the effects of background radiation. The revised analysis shows that the proposed diagnostic system can achieve the required accuracy of 10% for temperatures up to 20 keV at low plasma background radiation, whereas under high background radiation this accuracy can only be maintained up to 1 keV. To achieve the 10% accuracy across the full temperature range while preserving the current diagnostic configuration, either the energy of the probing laser pulse must be increased by a factor of 2–4 or the least electron density must be raised to $6\times10^{19}$ m$^{-3}$.

## 1. Introduction

A diagnostic system based on a 1064 nm Nd:YAG laser and filter polychromators has been proposed for measuring electron temperature in the ITER plasma core using the Thomson scattering technique [1–6]. In the adopted diagnostic configuration, the laser beam probes the plasma approximately horizontally along the tokamak major radius, while scattered radiation is collected from multiple observation points at scattering angles between 130° and 160° relative to the incident beam direction. Under these conditions, the Thomson backscattering spectra are substantially broadened, reducing their spectral intensity against the strong plasma radiation background. To compensate for this reduction, high-power lasers with short pulse width are employed to increase the intensity of the scattered signal. This diagnostic configuration presents particular challenges for measuring electron temperatures above 20 keV, as the Thomson scattering spectra become increasingly broadened and extend into the ultraviolet region. The ITER Thomson scattering diagnostic is designed with a Nd:YAG laser delivering 4 J pulses of 4 ns width [5], enabling electron temperature measurements up to 40 keV with an accuracy better than 10% over the density range $3\times10^{19}$–$3\times10^{20}$ m$^{-3}$. Scattered radiation from different plasma locations is analyzed using multichannel polychromators equipped with interference filters and avalanche photodiodes (APDs) in each spectral channel. The feasibility of meeting these performance requirements was demonstrated in [6] through simulations of Thomson scattering signals, expressed in terms of the number of photoelectrons generated in the APDs, and assessment of the accuracy of electron temperature measurements in the plasma centre at an electron density of $3\times10^{19}$ m$^{-3}$. The simulations accounted for the transmission of the optical system, the quantum efficiency of the APDs, and the contribution of plasma background radiation. The background radiation level was estimated for a steady-state ITER discharge with Q=10, taking into account plasma bremsstrahlung and line emission from the divertor being reflected from the tokamak walls [6]. In the present paper, it is shown that the number of photoelectrons reported in [6] was significantly overestimated, particularly in the short-wavelength region, because the calculations were based on energy scattering spectra rather than photon-number spectra. As a result, the predicted accuracy of electron temperature measurements was also overestimated. The revised analysis demonstrates that, to maintain the required measurement accuracy within the adopted diagnostic configuration, either the probing laser pulse energy must be increased by a factor of 2–4 or the minimum plasma density at which the required accuracy can be achieved must be raised.

## 2. Estimation of the number of photoelectrons in Thomson scattering signals

To estimate electron temperature measurement errors, Thomson scattering and background radiation signals in the spectral channels of a Thomson scattering diagnostic are typically expressed in terms of the number of detected photoelectrons, $N_{pe}$. This quantity is determined from the number of photons incident on the detector during a laser pulse, taking into account the detector quantum efficiency and the transmission of the optical system. The number of photoelectrons in an APD determines the variance of the signal noise, given by $\sigma^2=F*N_{pe}$, where $F$ is the detector excess noise factor. The measured signals and their associated variances are then used to determine the electron temperature, electron density, and their uncertainties by means of the least squares regression (LSR). The number of photoelectrons generated in the APDs of eight spectral channels of a polychromator collecting Thomson-scattered and background radiation from the ITER plasma was simulated in [6]. To obtain a reliable estimate of the temperature measurement uncertainty, the temperature errors were evaluated from the temperature variations in repeated calculations performed with randomly generated scattering and background signals obeying normal distributions. The simulation results indicated that, under the assumed discharge conditions and average divertor load, the diagnostic system could achieve the required temperature measurement accuracy of 10% for electron temperatures up to 40 keV. The same accuracy was predicted only up to 20 keV when the background radiation is increased by a factor of 5. To extend the measurement range to higher temperatures while maintaining the required accuracy, [6] proposed the use of an additional probing laser operating at a wavelength of 1320 nm, thereby shifting the peak of the Thomson scattering spectrum from the ultraviolet region into the spectral range covered by the polychromator.

However, these conclusions regarding the feasibility of measuring high electron temperatures in ITER plasmas were based on incorrect estimates of the number of photoelectrons in the polychromator spectral channels obtained using Eq. (1) of [6]. Inspection of this equation reveals that its left-hand and right-hand sides have inconsistent dimensions. This error can be corrected by dividing the right-hand side by the probing laser wavelength, $\lambda_0$. With this correction, and excluding the detector quantum efficiency $QE(\lambda)$ and the photon energy of the probing radiation, $hc/\lambda_0$, the right-hand side of the equation represents the spectral density function proportional to the energy of the scattered radiation, $E_{sc}$, collected at the detector input in the $i$-th spectral channel by an optical system with transmission $T\, T_s(\lambda)\, K_i(\lambda)$, see, for example, [7,8]:

$$E_{sc} = n_e L T \Omega E_L \int T_s(\lambda) K_i(\lambda) \sigma_{TS}(\lambda_0, \lambda, T_e, \theta) \frac{d\lambda}{\lambda_0} \qquad (1)$$

Here, $T$ is the total transmission of the optical system, $T_s(\lambda)$ is the spectral transmission of the collection optics, and $K_i(\lambda)$ is the spectral transmission of the $i$-th channel, determined by the corresponding interference filter. The Thomson scattering cross-section, $\sigma_{TS}$, is calculated for electrons with a relativistic Maxwellian distribution at temperature $T_e$ [8–10]. To determine the number of detected photoelectrons, the integral expression in Eq. (1) must be divided by the photon energy, $h\nu$, and multiplied by the detector quantum efficiency, $QE(\lambda)$. After straightforward transformations, the following expression is obtained:

$$N_{fei} = n_e LT\Omega N_L \int T_s(\lambda)K_i(\lambda)QE_i(\lambda)\sigma_{TS}(\lambda_0, \lambda, T_e, \theta)\frac{\lambda}{\lambda_0}\frac{d\lambda}{\lambda_0} \quad (2)$$

Here, $N_L = E_L/h\nu_0$ denotes the number of photons contained in a probing laser pulse. At the electron temperatures relevant to ITER measurements, the spectral distributions of the Thomson-scattered energy, $\sigma_{TS}$, and of the scattered photon number, $\sigma_{TS}\lambda/\lambda_0$, differ substantially, particularly at high temperatures, as illustrated in Fig. 1. The solid curves in Fig. 1 represent the Thomson scattering energy spectra, $\sigma_{TS}(\lambda, T_e)$, reproduced from Fig. 5 of [6]. The dash-dotted curves show the corresponding scattered-photon spectra, $\sigma_{TS}(\lambda, Te)\lambda/\lambda_0$. At elevated electron temperatures, these two kind spectra differ significantly, especially in the short-wavelength region. The figure also includes the layout of spectral channel adopted in [6].

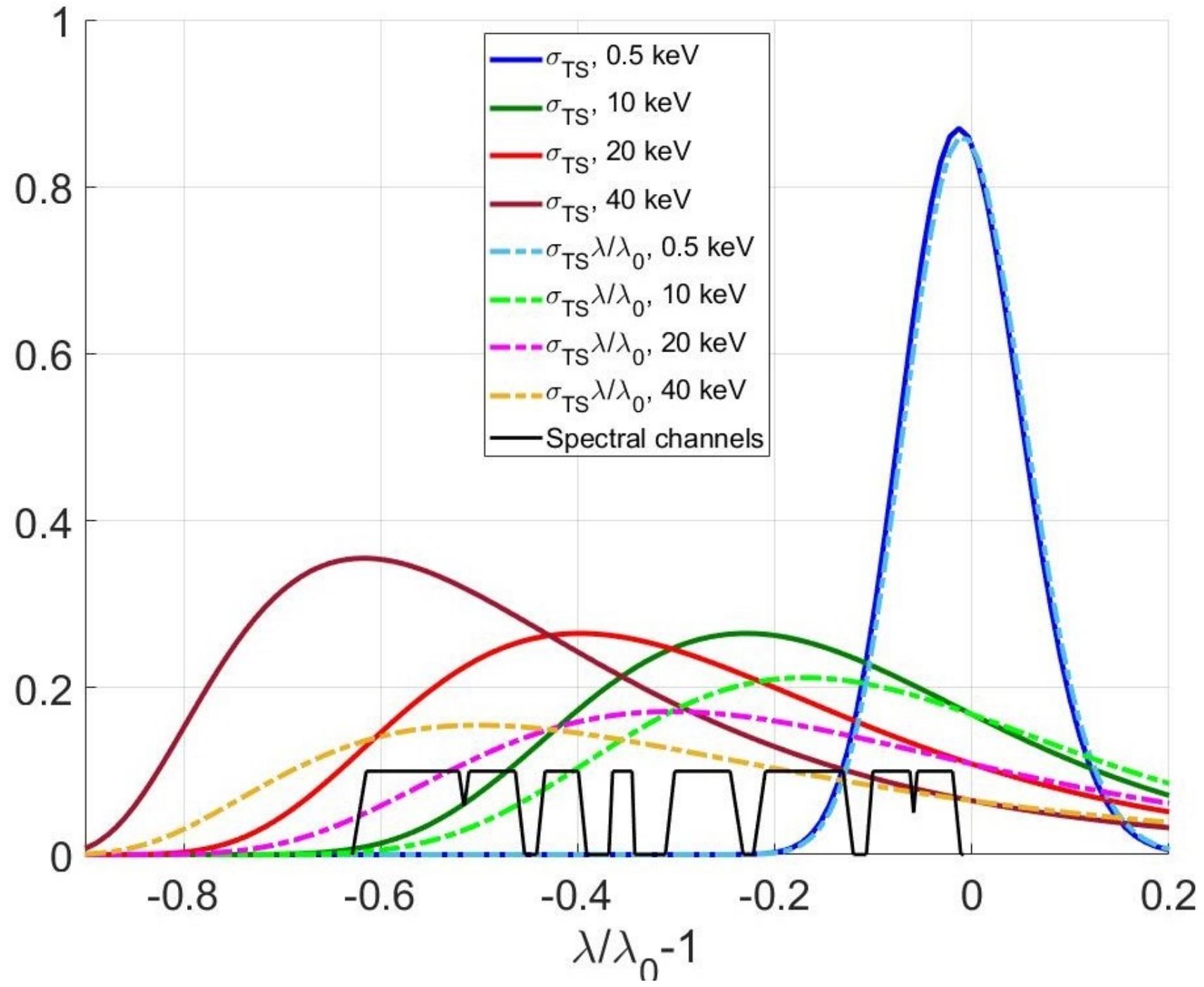

*Fig. 1 Spectra of scattered photon energy and photon numbers*

For measurements of Thomson-scattered radiation from the ITER plasma, two types of silicon photodetectors with a 3 mm entrance aperture are planned to be used: the Hamamatsu S8664-30K [11] for the short-wavelength region and the Excelitas C30956EH [12,13] for the visible and near-infrared regions of the scattering spectrum. In [6], the quantum efficiencies *QE* of these detectors were calculated from the spectral responsivities $R_D$ reported in [11–13] using the known relationship [14]:

$$QE = \frac{R_D h\nu}{eM_0} \quad (3)$$

Here, $e$ is the elementary charge, $h\nu$ is the photon energy, and $M_0$ is the avalanche gain of the detector. The gain was assumed to be independent of the wavelength of the detected radiation and was taken to be 50 for the S8664 detector and 75 for the C30956EH detector. The quantum efficiencies calculated using Eq. (3) are in full agreement with those reported in [6] and are presented in Fig. 2. The figure also shows the spectral transmission of the optical system used to collect scattered radiation from the plasma, together with the positions of the polychromator spectral channels, both reproduced from [6].

In [6], the number of photoelectrons in each spectral channel was calculated by integrating the Thomson scattering energy spectrum, $\sigma_{TS}(\lambda, T_e)$ over wavelength, weighted by the calculated detector quantum efficiency and the spectral transmission of the optical system.

$$N_{fei} = n_e LT\Omega N_L \int T_s(\lambda)K_i(\lambda)QE_i(\lambda)\sigma_{TS}(\lambda_0, \lambda, T_e, \theta)\frac{d\lambda}{\lambda_0} \quad (4)$$

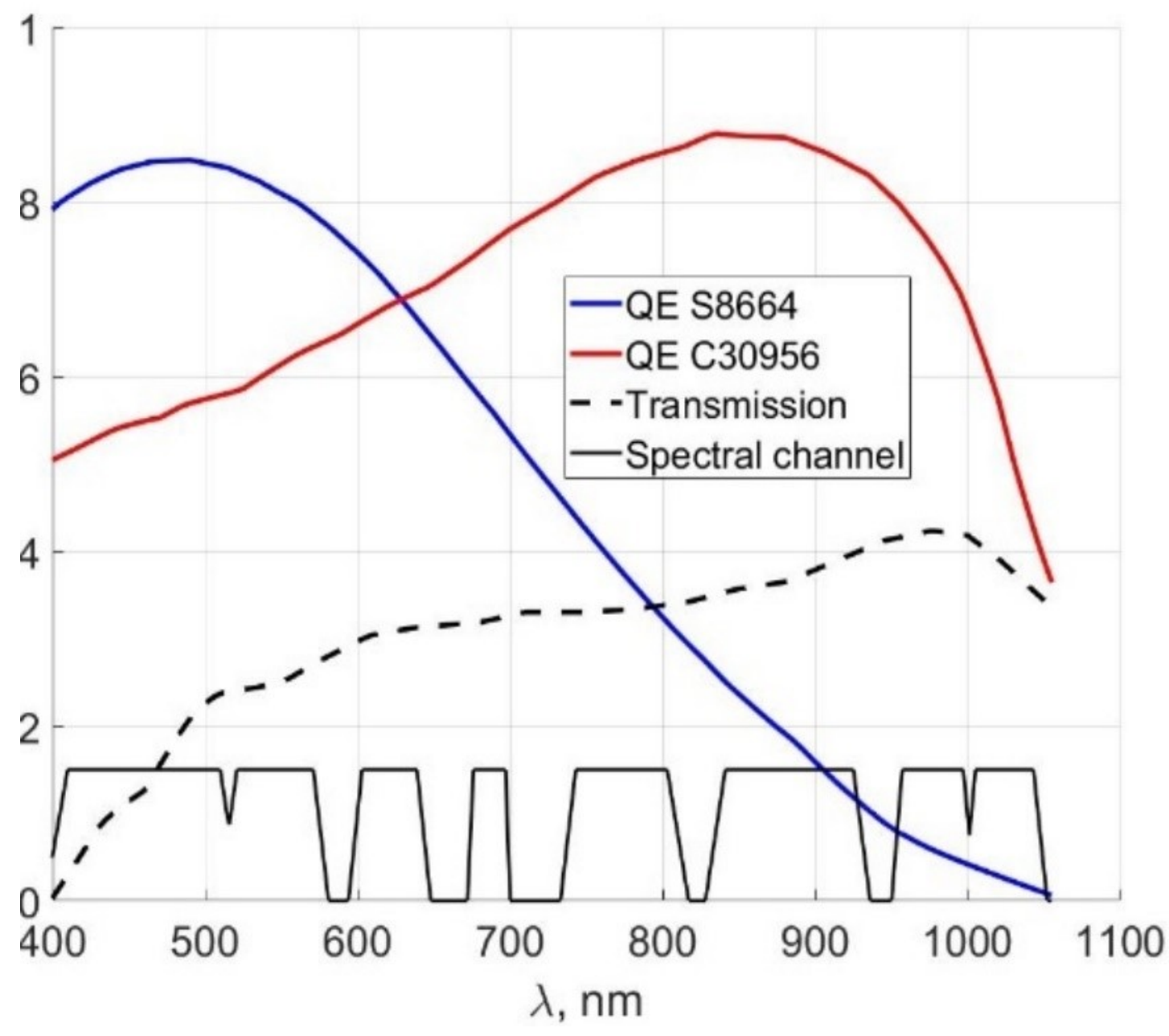

*Fig. 2 Calculated quantum efficiency of APDs and transmission of the optical system*

The use of this expression leads to an overestimation of the number of photoelectrons relative to the correct formulation given by Eq. (2), owing to the omission of the factor $\lambda/\lambda 0$ in the integrand. As a result, the electron temperature measurement errors reported in [6], particularly at high temperatures, are significantly underestimated.

## 3. Estimation of electron temperature measurement errors in ITER plasma

Transmission characteristics $K_i(\lambda)$ of the interference filters in the spectral channels are presented only schematically in [6] and they cannot be use for direct reassessment of the temperature measurement errors using Eq. (2). For this reason, the comparison was performed using an alternative approach.

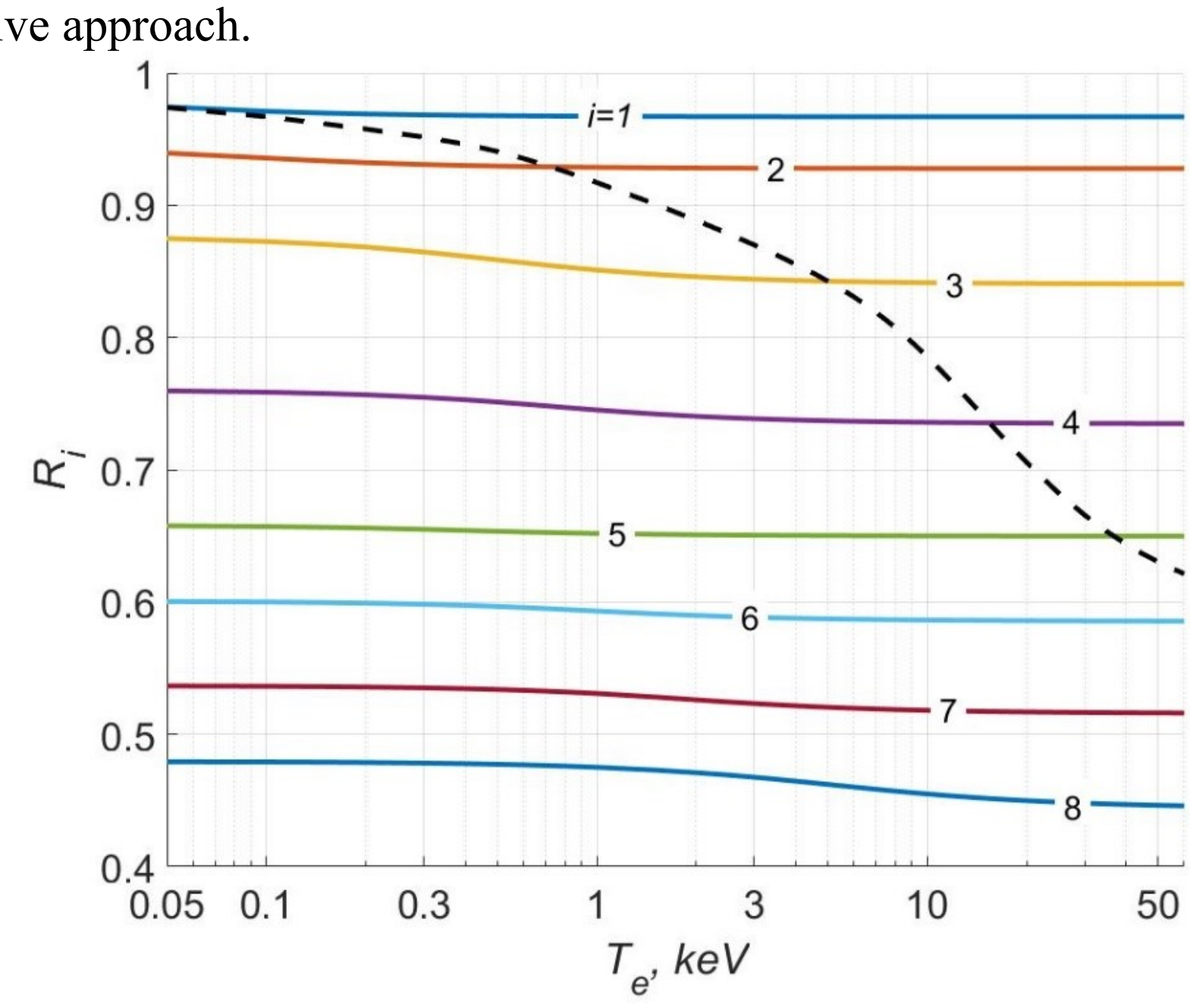

*Fig. 3 Ratio of photoelectron numbers $R_i$*

In the first stage, the numbers of photoelectrons in the spectral channels were calculated using the incorrect expression, Eq. (4), together with the polychromator optical transmission function $T_s(\lambda)$ shown in Fig. 2. At each wavelength, the value of $QE_i(\lambda)$ was taken as the highest quantum efficiencies of the two photodetectors. The channel filter transmission functions, $K_i(\lambda)$, were specified within the channel spectral ranges indicated in Fig. 2 so that the calculated temperature dependences of the photoelectron signals closely reproduced the results in [6].

In the second stage, the calculations were repeated using the correct expression, Eq. (2). For each spectral channel, correction factors $R_i$ were then determined as the ratios of the numbers of photoelectrons calculated using Eqs. (2) and (4). These ratios are stable in regards with variations of $K_i(\lambda)$ and plotted in Fig. 3 as functions of electron temperature. The dashed curve represents the corresponding ratio of the total number of photoelectrons summed over all channels. At high electron temperatures, the total numbers of photoelectrons predicted by Eqs. (4) and (2) are 2.1×104 and 1.3×104, respectively.

The calculated correction factors exhibit only a weak dependence on electron temperature and therefore were used to determine corrected photoelectron numbers in the spectral channels from the data presented in Fig. 4a of [6]. The corresponding plasma background signals were taken from the same figure. The electron temperature measurement errors were evaluated employing these data with the use of a formula from [15, 16] which is applied in [6].

The relative temperature errors calculated from the original data of [6] are in good agreement with the results reported in that paper. In Fig. 4, the solid and dashed red curves represent the relative temperature errors presented in [6] for two different levels of plasma background radiation. The corresponding errors recalculated in the present study from the digitized photoelectron data using Eq. (4) are shown by the magenta curves. The small discrepancies between the red and magenta curves are attributable to uncertainties introduced during the digitization of the graphical data.

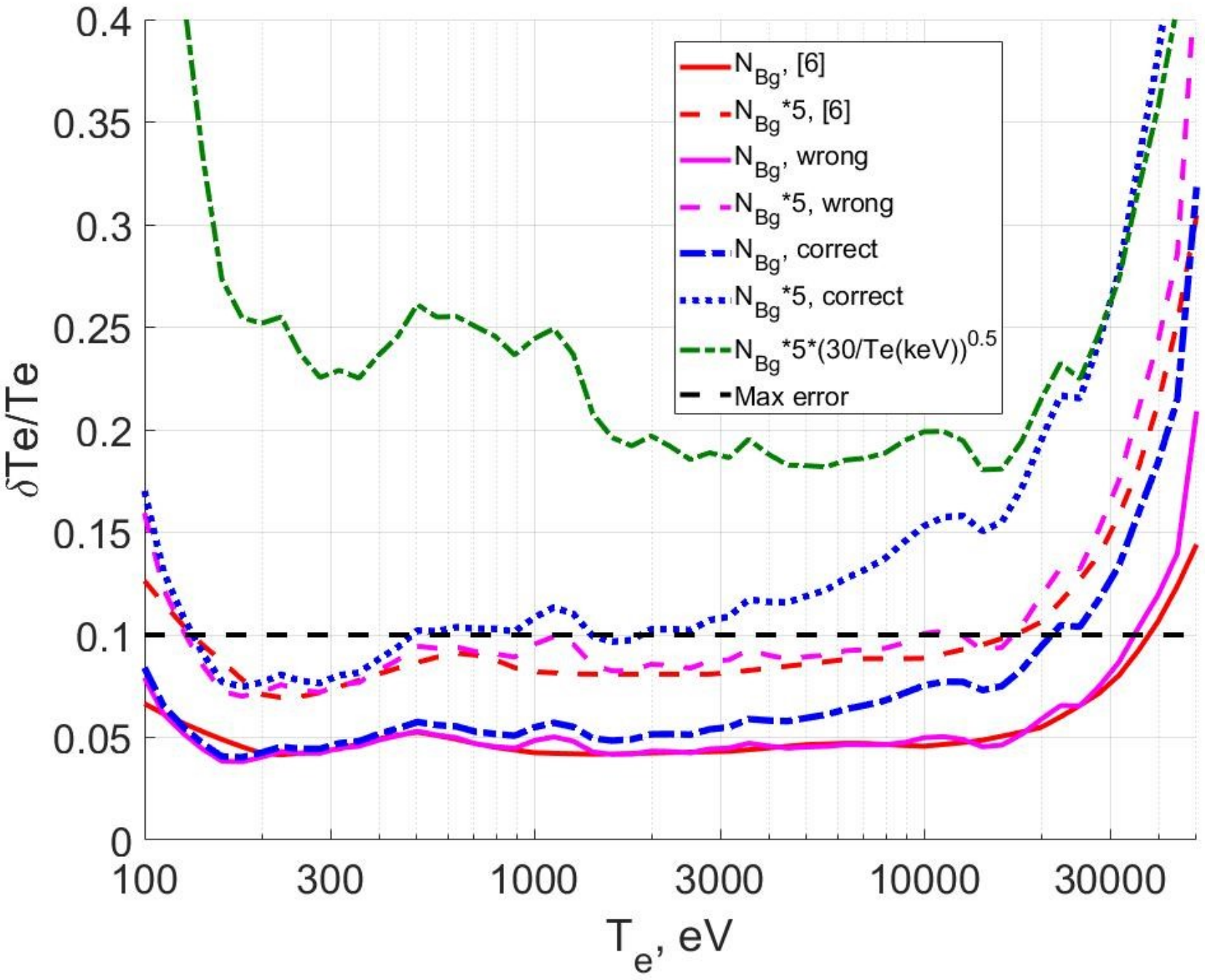

*Fig. 4 Accuracy of electron temperature measurements*

The electron temperature measurement errors obtained after correcting the photoelectron numbers using the coefficients Ri are shown in Fig. 4 by the blue curves for two levels of plasma background radiation. At the lower background level, deviations from the results reported in [6] become noticeable at temperatures above approximately 1 keV. At Te = 40 keV, the corrected measurement error approaches 20%, nearly twice the required limit of 10%. For the higher background level, the measurement error reaches 10% at an electron temperature of approximately 0.5 keV and continues to increase, reaching about 40% at 40 keV.

These estimates were obtained under the assumption that the background radiation levels in all spectral channels remain unchanged over the entire temperature range from 100 eV to 60 keV. Such an assumption is unlikely to be realistic. If the background levels presented in [6] were evaluated for plasma scenarios with high central temperatures, they may increase substantially at lower temperatures owing to the increase in bremsstrahlung intensity, which scales approximately as $T_e^{-1/2}$. This effect would lead to larger temperature measurement errors in the low-temperature region. To assess its impact, it was assumed that the background radiation level varies with electron temperature as $(30/T_e(keV))^{1/2}$. Under this assumption, the predicted measurement error is approximately 20% over the entire temperature range, as indicated by the green dash-dotted curve in Fig. 4.

## 4. Influence of avalanche photodiode characteristics on the Estimation of Photoelectron Yield and Electron Temperature Measurement Errors

In [6], the quantum efficiency of the avalanche photodiodes was determined using Eq. (3) under the assumption that the avalanche gain is independent of wavelength. Such an assumption is generally valid for photomultiplier tubes, where electron multiplication occurs through successive acceleration stages in vacuum between dynodes. However, it is not applicable to semiconductor avalanche photodiodes, as demonstrated by the difference between the calculated quantum efficiency of the Hamamatsu S8664 APD and its manufacturer's specifications, as presented in Fig. 5a with the dashed and solid blue curves. The red curve shows the spectral responsivity provided by the manufacturer for an avalanche gain of $M_0$ = 50, specified at a wavelength of 580 nm. The wavelength dependence of the avalanche gain, shown by the green dotted curve in Fig. 5a, was derived from the manufacturer's data using the relationship between spectral responsivity expressed in A/W and quantum efficiency of APD [14]:

$$M = \frac{R_D\, h\nu}{QE\; e} \quad (5)$$

An incorrect determination of the quantum efficiency of the S8664 APD in [6] results in an overestimation of the number of photoelectrons by approximately 5% in the spectral region $\lambda$<620 nm, where this detector is applied in the calculations.

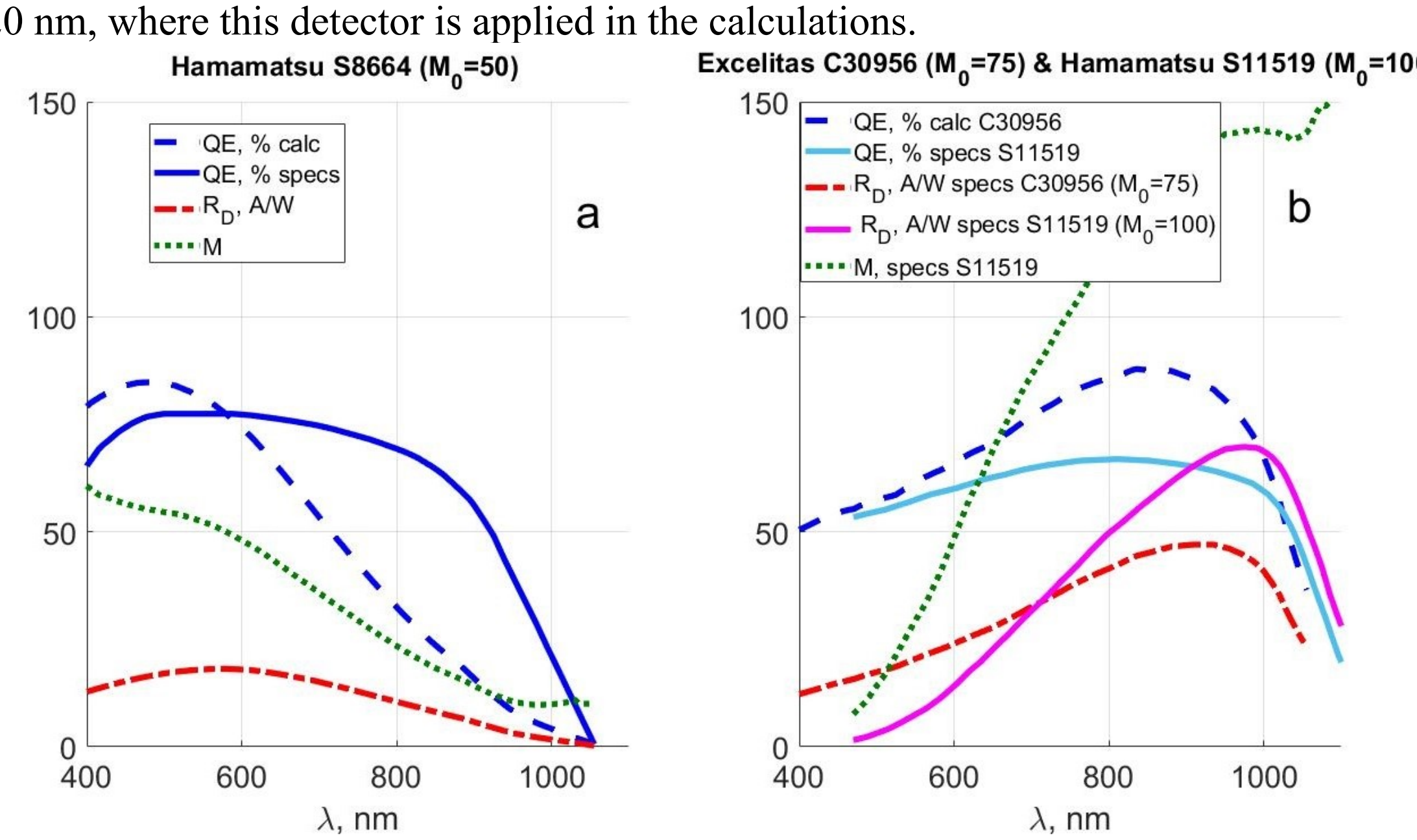

*Fig. 5 Spectral characteristics of APDs*

The quantum efficiency of the C30956 detector is not specified by the manufacturer; therefore, its spectral dependence calculated in [6] is compared in Fig. 5b with the quantum efficiency of another detector, the Hamamatsu S11519, as provided by the manufacturer. The calculated quantum efficiency of the C30956, shown by the blue dashed curve, is significantly higher than that of the S11519, shown by the solid blue curve. At the same time, the spectral responsivities of these detectors, shown by the dotted red and solid magenta curves, are approximately consistent when accounting for the difference in avalanche gain, 75 and 100, respectively.

Based on this comparison, it can be inferred that the actual quantum efficiency of the C30956 detector is lower than the calculated values, and consequently the number of photoelectrons obtained for this detector may be overestimated by up to 10%. These corrections do not lead to a significant increase in the estimated electron temperature measurement error. However, for the proper design of the detection system, it is necessary to determine the quantum efficiency of the C30956 detector experimentally in order to accurately estimate the number of photoelectrons and define its spectral range of applicability.

## 5. Conclusion

The presented analysis demonstrates the feasibility of measuring the central electron temperature in ITER plasma using a Thomson scattering diagnostic system in the adopted configuration at an electron density of $3\times10^{19}$ $m^{-3}$ at the required 10% accuracy for temperatures up to 20 keV under low plasma light background radiation conditions and up to 1 keV under high plasma light background conditions. When the increase in bremsstrahlung background at lower electron temperatures is taken into account, the measurement error may exceed 10% across the entire temperature range. To maintain the required accuracy, the laser pulse energy would need to be increased by a factor of 2–4 depending on the background radiation level. However, this approach is impractical because optical components can be very likely damaged by the intensive laser beam while preserving the short laser pulse width of 4 ns. The use of an additional laser with a probing wavelength of 1320 nm will not be able to provide the required accuracy, since its probing power remains an order of magnitude lower than that of routine lasers emitting at 1064 nm [17]. Alternatively, to achieve 10% accuracy while retaining the existing Thomson scattering diagnostic configuration, the minimum electron density threshold must be increased to above $6\times10^{19}$ $m^{-3}$.

## Acknowledgements

This work is carried out under a collaboration agreement between the Ioffe Institute and the Institute of Plasma Physics in Hefei. The work was supported by Ioffe Institute Contract FFUG-2024-0028 (Section 2), by the National Natural Science Foundation of China under Grant No. 12135015 (Section 3) and PIFI of the Chinese Academy of Sciences under Grant No.2025PVA0154 (Section 4 and writing the paper).